\documentclass[10pt,conference]{IEEEtran}
\usepackage{cite}
\usepackage{amsmath,amssymb,amsfonts}
\usepackage{algorithmic}
\usepackage{subcaption}
\usepackage{graphicx}
\usepackage{textcomp}
\usepackage{xcolor}
\usepackage{booktabs}
\usepackage{enumitem}
\usepackage{tcolorbox}
\usepackage[hidelinks]{hyperref}
\def\BibTeX{{\rm B\kern-.05em{\sc i\kern-.025em b}\kern-.08em
    T\kern-.1667em\lower.7ex\hbox{E}\kern-.125emX}}

\newcommand{\RqOne}{\textbf{(RQ1):} \emph{How prevalent are links in source code comments?}}

\newcommand{\RqTwo}{\textbf{(RQ2):} \emph{What kind of link targets are referenced in source code comments?}}

\newcommand{\RqThree}{\textbf{(RQ3):} \emph{What purpose do links in source code serve?}}

\newcommand{\RqFour}{\textbf{(RQ4):} \emph{How do links in source code comments evolve?}}

\newcommand{\RqFive}{\textbf{(RQ5):} \emph{How frequently do link targets referenced in source code comments change?}}

\newcommand{\RqSix}{\textbf{(RQ6):} \emph{How many links in source code comments are dead?}}

\newcommand{\RqSeven}{\textbf{(RQ7):} \emph{Can we fix dead links in source code comments?}}

\begin{document}

\title{9.6 Million Links in Source Code Comments:\\Purpose, Evolution, and Decay}


\author{\IEEEauthorblockN{Hideaki Hata\IEEEauthorrefmark{1},
Christoph Treude\IEEEauthorrefmark{2},
Raula Gaikovina Kula\IEEEauthorrefmark{1} and
Takashi Ishio\IEEEauthorrefmark{1}}
\IEEEauthorblockA{\IEEEauthorrefmark{1}Nara Institute of Science and Technology\\
\{hata, raula-k, ishio\}@is.naist.jp}
\IEEEauthorblockA{\IEEEauthorrefmark{2}University of Adelaide\\
christoph.treude@adelaide.edu.au}
}

\maketitle

\begin{abstract}
Links are an essential feature of the World Wide Web, and source code repositories are no exception. However, despite their many undisputed benefits, links can suffer from decay, insufficient versioning, and lack of bidirectional traceability. In this paper, we investigate the role of links contained in source code comments from these perspectives. We conducted a large-scale study of around 9.6 million links to establish their prevalence, and we used a mixed-methods approach to identify the links' targets, purposes, decay, and evolutionary aspects. We found that links are prevalent in source code repositories, that licenses, software homepages, and specifications are common types of link targets, and that links are often included to provide metadata or attribution. Links are rarely updated, but many link targets evolve. Almost 10\% of the links included in source code comments are dead. We then submitted a batch of link-fixing pull requests to open source software repositories, resulting in most of our fixes being merged successfully. Our findings indicate that links in source code comments can indeed be fragile, and our work opens up avenues for future work to address these problems.
\end{abstract}


\section{Introduction and Motivation}

When Ted Nelson started Project Xanadu\footnote{\url{http://www.xanadu.com/}} in 1960, he envisioned ``an entire form of literature where links do not break as versions change; where documents may be closely compared side by side and closely annotated; where it is possible to see the origins of every quotation; and in which there is a valid copyright system---a literary, legal and business arrangement---for friction-less, non-negotiated quotation at any time and in any amount''~\cite{nelson1999xanalogical}. 
Links were supposed to be visible and could be followed from all endpoints, with permission to link to a document explicitly granted by the act of publication~\cite{aiello2018web}. 
Decades later, Nelson witnessed the birth of the World Wide Web, which in his words ``trivialized this original Xanadu model, vastly but incorrectly simplifying these problems to a world of fragile ever-breaking one-way links, with no recognition of change or copyright, and no support for multiple versions or principled re-use''~\cite{nelson1999xanalogical}. 
As predicted by Nelson, the Internet and its implementation of links have afforded us countless opportunities since, but also experienced issues such as link decay~\cite{markwell2002broken, klein2014scholarly}, digital plagiarism~\cite{barrie2000digital}, and the need to rely on external services to keep historical copies of web content~\cite{murphy2007take}.

In this work, we investigate the role of links contained in source code comments from the perspective of these opportunities and challenges: what purposes do they serve, how do they and their targets evolve, and how often do they break? 
The significance of this work is closely related to software documentation~\cite{8094446} and self-admitted technical debt~\cite{potdar2014exploratory}. To improve documentation and mitigate potential issues, it is important to understand developers' typical knowledge sharing activities by referencing external sources, and to investigate link decay as a potential problem.

Our work is related to and inspired by recent research on source code comments in terms of documentation, traceability, licensing, and attribution. 
For example, source code comments have been found to document technical debt~\cite{potdar2014exploratory} and to support articulation work~\cite{storey2008todo}. 
They are fragile with respect to identifier renaming, i.e., traceability between comments and code is easily lost~\cite{Ratol2017}.
Source code comments located at the beginning of a file often include a text or a link indicating the copyright and license information of the file~\cite{German:2010:SMA:1858996.1859088}. 
These comments are updated during the evolution of a product by the copyright holders~\cite{Wu2016}. 
Links in source code comments are sometimes used for attribution when source code has been taken from elsewhere---however, the vast majority of code snippets is copied without attribution~\cite{baltes2018sotorrent, baltes2018usage}. 
Despite these research efforts, to the best of our knowledge, the role of links in source code comments has not been studied comprehensively so far.

To fill this gap, in this paper, we first lay the foundation for understanding the role of links in source code comments by collecting 9,654,702 links from source code comments in 25,925 Git repositories. 
Our parser is able to extract comments from source code written in 7 programming languages. 
We find that links in source code comments are common: more than 80\% of the repositories in our study contained at least one link. 
Through a qualitative study of a stratified sample of 1,146 links, we establish the kinds of link targets that are referenced in source code comments.
To understand how links are used to indicate issues related to attribution, technical debt, copyright, and licensing, our qualitative study also uncovers the various purposes for including links in source code comments. 
We find that licenses, software homepages, and specifications are among the most prevalent types of link targets, and that links are often used to provide metadata or attribution.

Link decay has the potential of making documentation in source code comments fragile and buggy.
We investigate this issue from two perspectives: we analyze the evolution of the links in the repositories' commit histories and we examine how often link targets referenced in source code comments change. 
We find that links are rarely updated, but their targets evolve, in almost 10\% of all cases leading to dead links.
We then submit fixes to a subset of these broken links as pull requests, most of which were successfully merged by the maintainers of the corresponding open source projects.

In summary, this paper's contributions are three-fold:

\begin{itemize}
\item a large-scale and comprehensive study of around 9.6 million links to establish the prevalence of links in source code comments,
\item a mixed-methods study to identify targets, purposes, and evolutionary aspects of links in source code comments, and
\item an analysis of the extent to which links in source code comments are affected by link decay, with all nine link-fixing pull requests submitted to active open source projects already merged by the projects' maintainers. 
\end{itemize}



\section{Research Method}
\label{sec:rm}
In this section, we present our research questions and data collection methodology, and we introduce the data contained in our online appendix.

\subsection{Research Questions}
The main goal of the study is to gain insights into the purposes, evolution and decay of links in source code comments.
Based on this goal, we constructed seven research questions to guide our study.
We now present each of these questions, along with the motivation for each.

\begin{description}
\item \RqOne~
\end{description}
The motivation of \textbf{RQ1} is to understand whether the use of links in source code is a common practice in the wild. 
Furthermore, we would like to quantitatively explore the distribution, diversity, and spread of these links across different types of software projects.

\begin{description}
\item \RqTwo~
\item \RqThree~
\end{description}
\textbf{RQ2} and \textbf{RQ3} require a deeper analysis of the repositories, where we would like to understand the nature and purpose that the links serve.
The key motivation for \textbf{RQ2} is to identify the types of link targets that developers are likely to refer to in source code comments.
Furthermore, we would like to characterize the most common types of linked domains.
The key motivation for \textbf{RQ3} is to determine the reasons why developers use links.

\begin{description}
\item \RqFour~
\item \RqFive~
\item \RqSix~
\end{description}
\textbf{RQ4}, \textbf{RQ5}, and \textbf{RQ6} investigate the phenomenon of links in source code comments from an evolutionary and maintenance standpoint. We would like to understand whether developers are updating or maintaining these links after introducing them to the source code, whether the targets evolve, and how many of the links are affected by link decay.

\begin{description}
\item \RqSeven~
\end{description}
Our aim for \textbf{RQ7} is two-fold.
First, we would like to show that dead links can be restored and fixed.
Second, the results can serve as a validation that developers care about the maintenance of links in their code.

\subsection{Data Collection}
We now describe our methods for repository preparation, comment extraction, and link identification.

\textbf{Repository preparation.}
In this work, we analyzed \textit{active} software development repositories on GitHub written in \textit{common} programming languages. As common programming languages, we selected seven languages: C, C++, Java, JavaScript, Python, PHP, and Ruby. These languages have been ranked consistently in the top 10 languages on GitHub from 2008 to 2017 (based on the number of repositories from 2008 to 2015~\cite{github2015blog}, the number of pull requests from 2014 to 2017~\cite{githut}, and the number of pull requests in 2017~\cite{github2017octoverse}).

Using the GHTorrent dataset\footnote{MySQL database dump 2018-04-01 from \url{http://ghtorrent.org/downloads.html}.}~\cite{Gousios:2013:GDT:2487085.2487132}, we collected active repositories for the seven languages using the following criteria: (i) having more than 500 commits in their entire history (the same threshold used in previous work~\cite{Aniche:2018:CSM:3238579.3238606}), and (ii) having at least 100 commits in the most active two years. We designed the second criterion to remove long-term less active repositories and short-term projects that have not been maintained for long (and may not be software development projects). For example, we were able to exclude \texttt{software-engineering-amsterdam/sea-of-ql}, which is a repository of a collaboration space for students in a particular university course, and was reported as a false positive of software project identification~\cite{Munaiah:2017:CGE:3147777.3147808}. We determine repositories' languages based on the GHTorrent information. 
Forked repositories are excluded if repositories are recorded in GHTorrent as forks of other repositories.

\begin{table}
\centering
\caption{Collected repositories and links}
\label{tab:projectlists}
\begin{tabular}{lrrr}
\toprule
& \textbf{\# candidate repo} & \textbf{\# obtained repo (\%)} & \textbf{\# links} \\
\midrule
C & 2,771 & 2,482 (90\%) & 1,602,156 \\
C++ & 3,563 & 3,211 (90\%) & 1,686,575 \\
Java & 4,995 & 4,472 (90\%) & 2,925,909 \\
JavaScript & 7,130 & 6,224 (87\%) & 1,533,219 \\
Python & 5,263 & 4,715 (90\%) & 413,020 \\
PHP & 3,279 & 2,827 (86\%) & 1,405525 \\
Ruby & 2,233 & 1,994 (89\%) & 88,298 \\
\midrule
\textbf{sum} & \textbf{29,234} & \textbf{25,925 (89\%)} & \textbf{9,654,702} \\
\bottomrule
\end{tabular}
\end{table}

With the above criteria, we prepared the candidate list of target repositories for the seven languages as shown in Table~\ref{tab:projectlists}. When we collected these candidate repositories (from May to June 2018), some repositories were not available because they had been deleted or made private. In total, we obtained more than 25,000 repositories, which is almost 90\% of the candidate repositories.

\textbf{Comment extraction.}
From each Git repository, we extract source files of the labeled language in the \texttt{HEAD} commit (the latest snapshot of a cloned repository).
For example, only \verb|.java| files are extracted from a Java repository.
To process source files, we employ ANTLR4 lexical analyzers for six languages other than Ruby because their grammar definitions are available in the official example repository.\footnote{\url{https://github.com/antlr/grammars-v4/}}
For Ruby, we use a standard library, Ripper parser.

We extract all single line comments (e.g., \texttt{//} in C) and multiline comments (\texttt{/* ... */}) according to the grammars.
In the case of Python, string literals (\texttt{''' ... '''}) are also regarded as comments because they include documents (known as \textit{docstring}).
In the case of PHP, both HTML comments and PHP code comments are extracted.

\textbf{Link identification.}
From the extracted comments, links are identified using the regular expression \texttt{/http$\backslash$S+/} (\textit{localhost} and IP addresses, which are mainly used for private addresses, are excluded) and validated with the Perl module \texttt{Data::Validate::URI}.
We identified a total of 9,654,702 links from the collected repositories as seen in Table~\ref{tab:projectlists}. All links are recorded with the information of the corresponding file, repository identifiers (pairs of account and repository names), commit hashes, and the line number where the surrounding comment starts.
Considering the number of repositories, we found that repositories written in C, C++, and Java tend to contain more links compared to repositories in Python and Ruby.


\subsection{Online Appendix}
\label{ssec:appendix}

Our online appendix contains our 9,654,702 links associated with the information of languages and comment location (GitHub links including account names, repository names, commit hashes, file paths, and line numbers). The appendix is available at \url{https://github.com/NAIST-SE/9.6MillionLinks}.

\section{Findings}
\label{sec:find}
In this section, we present our findings for each research question.

\subsection{Prevalence of Links (RQ1)}
\label{ssec:prevalence}

\begin{figure*}[t!]
    \centering
    \begin{subfigure}[t]{\columnwidth}       
        \centering
			\includegraphics[width=.9\linewidth]{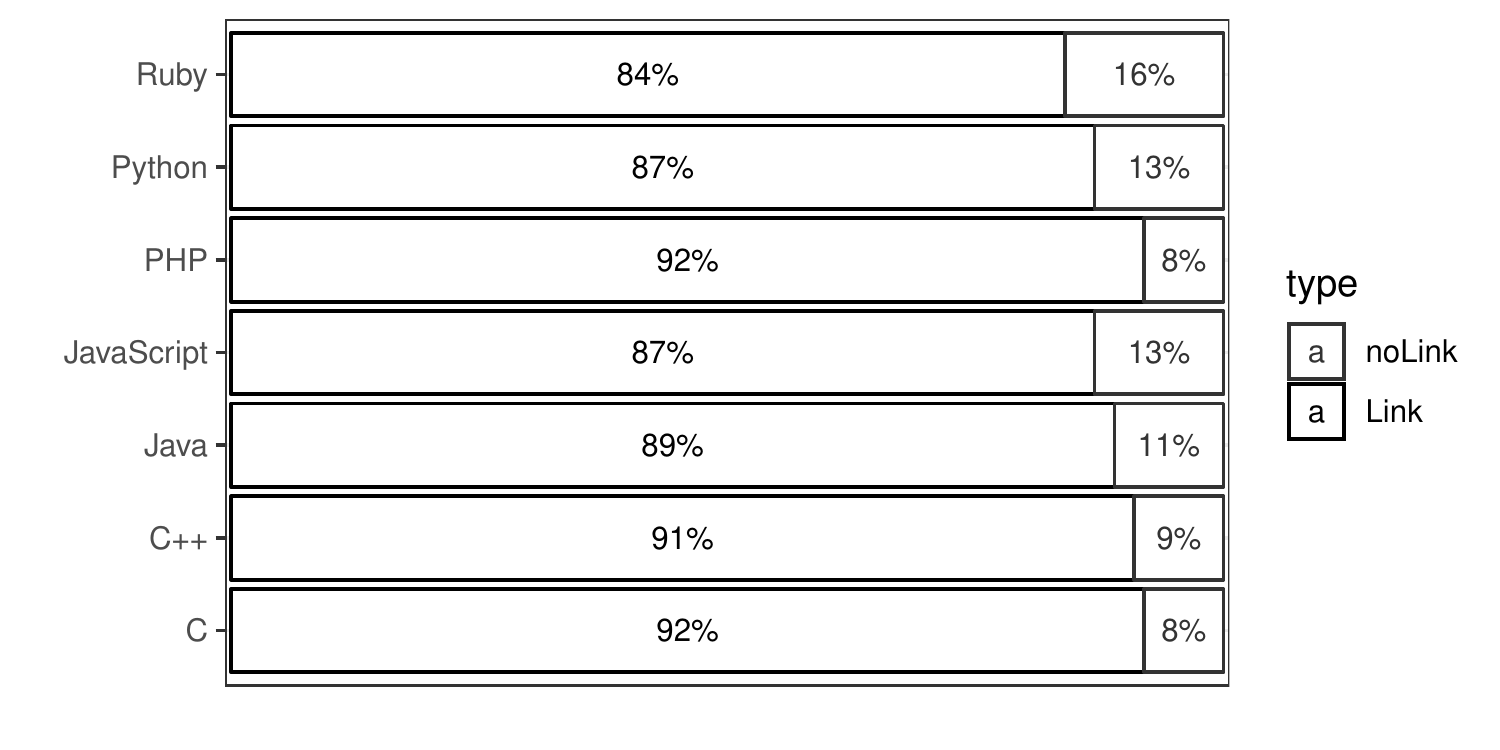}
		\caption{Ratio of repositories with links}
	\label{fig:RQ1a}
    \end{subfigure}
    \begin{subfigure} [t]{\columnwidth}      
        \centering
			\includegraphics[width=.9\linewidth]{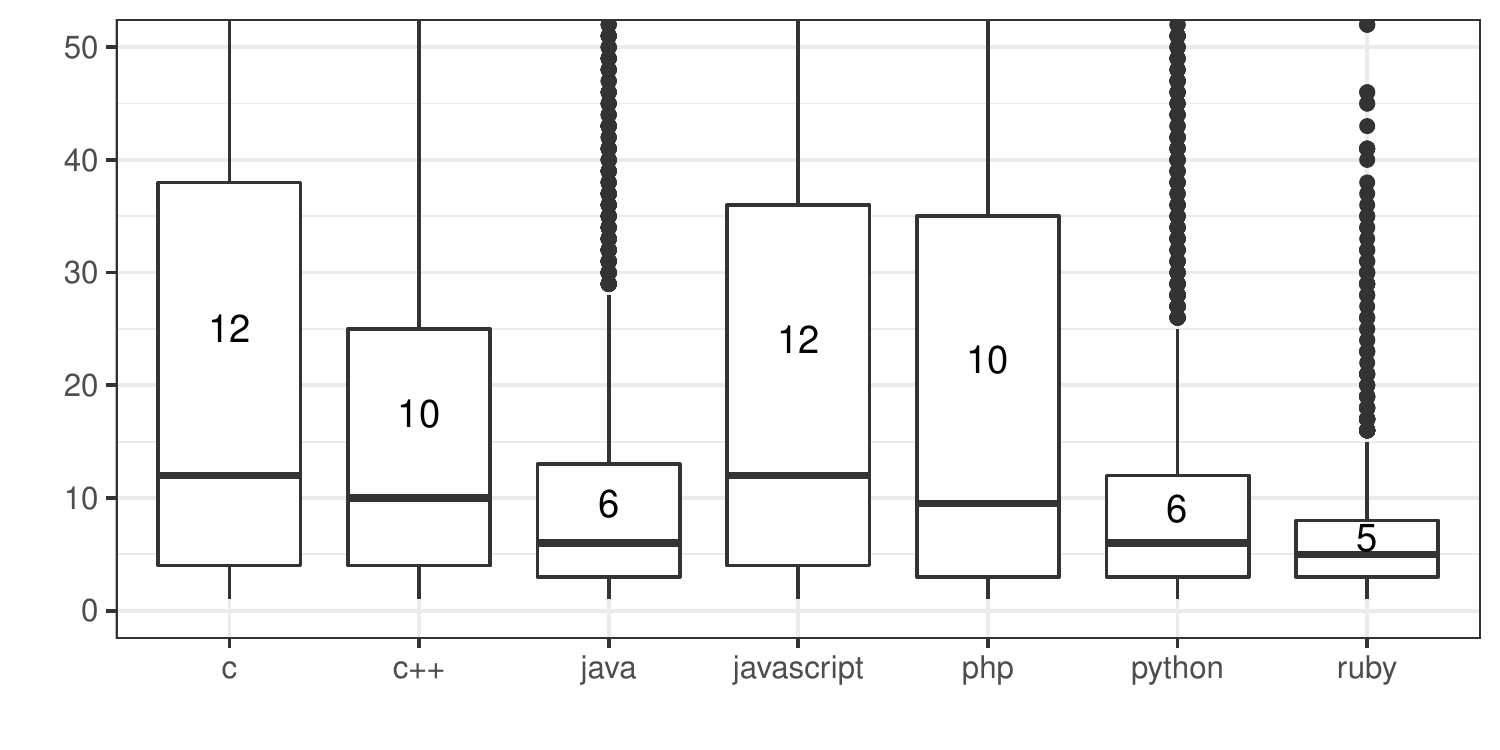}
		\caption{Distribution of number of different domains per repository}
	\label{fig:RQ1b}
    \end{subfigure}
    \begin{subfigure} [t]{\linewidth}      
        \centering
			\includegraphics[width=.9\linewidth]{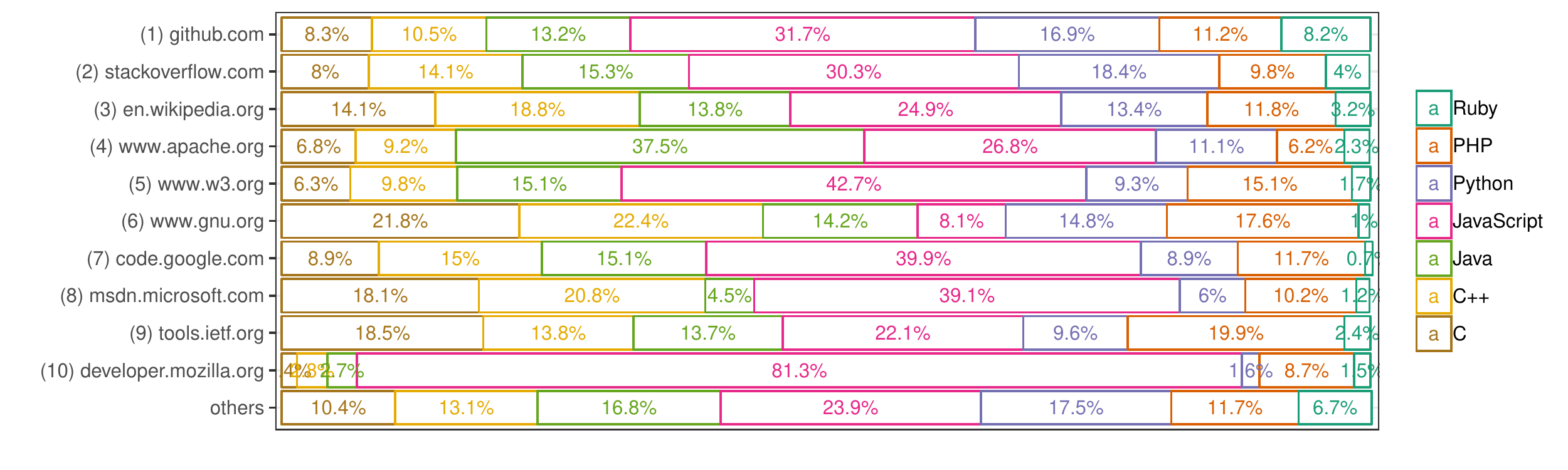}
		\caption{
		Proportion of repository languages shared by the top 10 most referenced domains.
		}
	\label{fig:RQ1c}
    \end{subfigure}
    \caption{Analysis of links by (a) languages, (b) domain diversity, and (c) top domains for RQ1}
    \label{fig:RQ1}
\end{figure*}
To understand the prevalence of links referenced in source code comments (\textbf{RQ1}), we conducted a quantitative analysis of our collected dataset in terms of link existence, domain diversity, and domain popularity.

\textbf{Link existence.}
Figure~\ref{fig:RQ1a} show the percentages of repositories that have at least one link in their source code comments. We see that, in every language, more than 80\% of the repositories contain links in source code comments. Especially for repositories written in C, C++, and PHP, more than 90\% of the repositories refer to external sources via links.

\textbf{Domain diversity.}
In the obtained 9,654,702 links, there are 57,039 distinct domains (Internet hostnames).
Figure~\ref{fig:RQ1b} shows the distribution of the number of distinct domains per repository, for repositories that have at least one link in their source code comments.
Median values are presented in the figure. We found that there is a diversity of links in a single repository even when summarized by their domains. Especially in repositories written in C, C++, JavaScript, and PHP, source code comments link to 10 or more different domains (median).

\textbf{Popular domains.}
Figure~\ref{fig:RQ1c} illustrates the proportion of languages shared by the top 10 most referenced domains.
Note that domain ranking is based on the number of repositories instead of the number of links. 
If links belonging to a domain appear in a small number of repositories, the domain will be low-ranked even if those repositories contain many links.

The \texttt{github.com} domain is the top referenced domain in our dataset. More than 14,000 repositories across seven languages referenced content on \texttt{github.com}. As we will describe in detail in Section~\ref{ssec:target}, such content includes \textit{software homepage}, \textit{code}, and \textit{profile of a GitHub contributor}.
However, we find in Section~\ref{ssec:decay} that many links to \texttt{github.com} are no longer available.
We also found many links to \texttt{code.google.com} (7th rank). 
Such content includes \textit{bug report}, \textit{software homepage}, and \textit{code}.
In a statistically representative sample of common domains (sampling described in Section~\ref{ssec:target}), two out of three links to \texttt{code.google.com} are redirected to \texttt{github.com}, and one links to \texttt{code.google.com/archive/}.

The \texttt{stackoverflow.com} domain is the second most referenced domain and has been linked to from 8,189 repositories.
As identified in previous work, Stack Overflow is widely used as a knowledge exchange platform between programmers~\cite{Treude:2011:PAA:1985793.1985907}, where programmers can obtain knowledge of good practices from code examples~\cite{Sillito:2012:MGC:2473496.2473558,7476684}, for example.
The large number of links to \texttt{stackoverflow.com} in source code comments can be another piece of evidence of developers' needs for knowledge acquisition from external resources. We study how code could be obsolete by not being updated when external sources change in Section~\ref{ssec:shortevol}.

The top domains differ by programming language: The \texttt{www.apache.org} domain is frequently linked from Java repositories, and the \texttt{www.gnu.org} domain is referenced from C and C++ repositories. Repositories written in JavaScript have many links to the Web-related domains of \texttt{www.w3.org} and \texttt{developer.mozilla.org}.

\begin{tcolorbox}
\textbf{Summary}: We revealed that links in source code comments are prevalent. In more than 80\% of the 25,925 active repositories written in seven common languages, there exists at least one link in each repository.
The top three most frequently referenced domains per repository are \texttt{github.com}, \texttt{stackoverflow.com}, and \texttt{en.wikipedia.com}.
\end{tcolorbox}

\subsection{Link Targets (RQ2)}
\label{ssec:target}

\begin{figure}
\centering
\includegraphics[width=\linewidth]{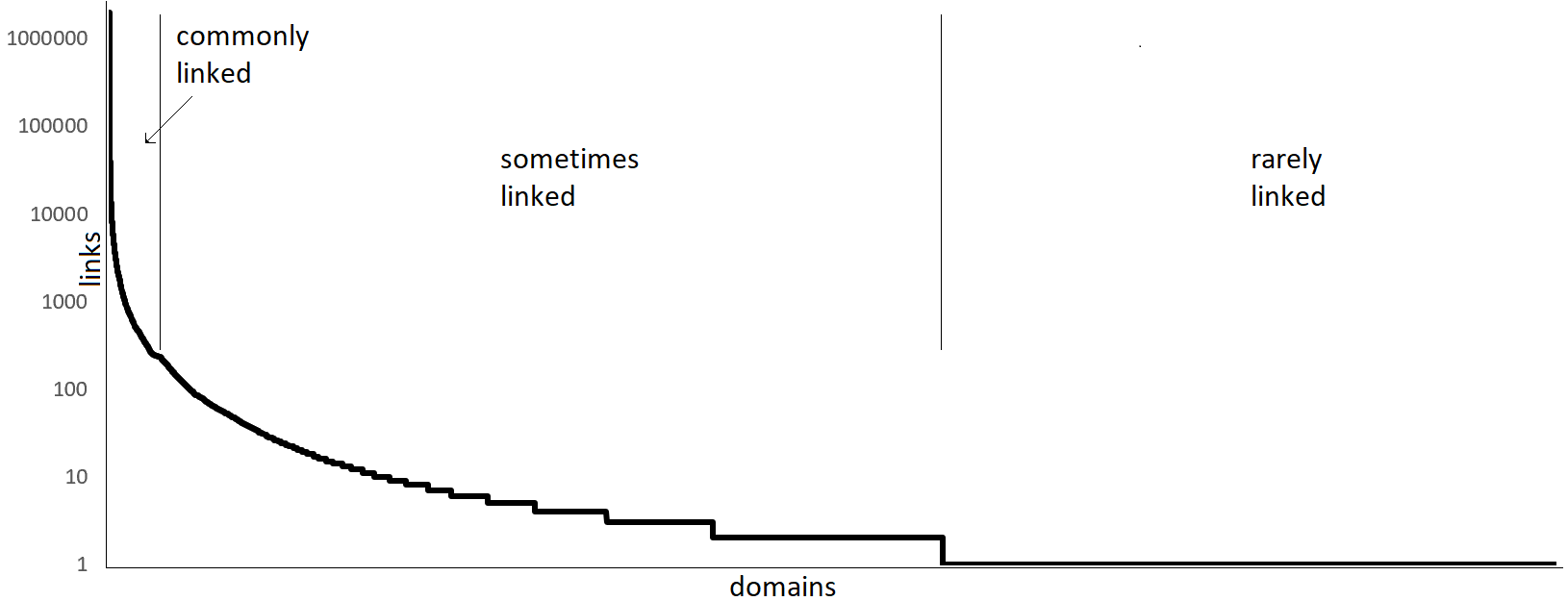}
\caption{Distribution of links per domain}
\label{fig:sample}
\end{figure}

To understand what kind of link targets are referenced in source code comments (\textbf{RQ2}), we conducted a qualitative study of a statistically representative and stratified sample of all links in our dataset. 

After an initial analysis of the link data, it quickly became obvious that some domains account for many links while other domains are rare. Based on this observation and to ensure diversity of our sample, we divided the data into three strata:

\begin{enumerate}
    \item links to \textit{commonly} linked domains,
    \item links to domains \textit{sometimes} linked, and
    \item links to \textit{rarely} linked domains.
\end{enumerate}

To decide on thresholds for distinguishing domains into those that are commonly, sometimes, and rarely linked, we conducted a visual analysis of the distribution of links per domain in our dataset. Figure~\ref{fig:sample} shows this distribution using a log scale. While content from the most commonly linked domain was linked more than a million times, many domains appeared in our dataset with a much lower frequency. We used the ``step'' in the distribution on the left-hand side of Figure~\ref{fig:sample} to distinguish between domains that are commonly linked and domains that are sometimes linked, with a cutoff frequency of 230. We consider domains which account for exactly one link in our dataset to be rarely linked.
Table~\ref{tab:sampling} shows the number of domains and the number of links in each strata. We then drew a statistically representative sample from each bucket. The required sample size was calculated so that our conclusions about the ratio of links with a specific characteristic would generalize to all links in the same bucket with a confidence level of 95\% and a confidence interval of 5.\footnote{\url{https://www.surveysystem.com/sscalc.htm}} The calculation of statistically significant sample sizes based on population size, confidence interval, and confidence level is well established (first published by Krejcie and Morgan in 1970~\cite{KrejcieMorgan1970}).

\begin{table}
\centering
\caption{Construction of the stratified sample}
\label{tab:sampling}
\begin{tabular}{l|rrrr}
\toprule
\textbf{strata} & \textbf{\# domains} & \textbf{\# links} & \textbf{\# links in sample} \\
\midrule
common & 2,013 & 9,128,444 & 384 \\
sometimes & 30,851 & 502,083 & 384 \\
rare & 24,175 & 24,175 & 378 \\
\midrule
\textbf{sum} & \textbf{57,039} & \textbf{9,654,702} & \textbf{1,146} \\
\bottomrule
\end{tabular}
\end{table}

The qualitative analysis was conducted in multiple iterations: in the first iteration, the first two authors independently coded 20 links from the sample, discussed a common coding guide, and tested this coding guide on another 20 links from the sample, refining the guide, merging codes, and adding codes which had been missed. The initial codes were informed by those used by Aniche et al.~\cite{Aniche2018} to categorize content posted on news aggregators, however, we found that their codes did not cover all types of link targets present in our dataset. In the second iteration, the four authors of this paper then independently coded another 30 links from the sample, using the coding guide designed by the first two authors. We then calculated the kappa agreement of this iteration between all four raters, for 30 cases and all 19 codes that emerged from the qualitative analysis.\footnote{Kappa agreement was calculated using \url{http://justusrandolph.net/kappa/}.} The kappa agreement was 0.81 or ``almost perfect''~\cite{viera2005understanding}. Based on this encouraging result, the remaining data was then coded by a single author.

The following list shows the 19 codes that emerged from our analysis along with a short description which was available in the coding guide:

\begin{itemize}
\item \textit{404}: link target does not exist (anymore) or cannot be accessed
\item \textit{licence}: licence of a software project
\item \textit{software homepage}: main web presence of a library or software project
\item \textit{specification}: anything that resembles a requirements document or a technical standard
\item \textit{organization homepage}: main web presence of an organization or company
\item \textit{other}: anything that does not fit the other codes, including if sign-in is required
\item \textit{tutorial or article}: technical article or tutorial, without commenting section (blog post otherwise)
\item \textit{API documentation}: documentation of an API element
\item \textit{blog post}: technical content with a commenting section
\item \textit{application}: interactive application (e.g., web application, online utility)
\item \textit{bug report}: bug report or issue in an online bug/issue tracker
\item \textit{research paper}: academic paper
\item \textit{personal homepage}: personal homepage of one individual
\item \textit{code}: a source code file
\item \textit{forum thread}: thread in a forum or entire forum
\item \textit{GitHub profile}: profile of a GitHub contributor
\item \textit{book content}: chapter/section of a book or entire book
\item \textit{Q\&A thread}: question-and-answer thread, but not Stack Overflow
\item \textit{Stack Overflow}: question-and-answer thread on Stack Overflow
\end{itemize}

\begin{table}
\centering
\caption{Frequency of link target types in our sample}
\label{tab:rq2results}
\begin{tabular}{l|r@{}rr@{}rr@{}r}
\toprule
 & \multicolumn{2}{c}{\textbf{common}} & \multicolumn{2}{c}{\textbf{sometimes}} & \multicolumn{2}{c}{\textbf{rare}} \\
\midrule
404                   & 27  & (7\%)       & 122 & (32\%)  & 138 & (37\%)  \\
license               & 208 & (54\%)      & 4   & (1\%)   & 1   & (0\%)   \\
software homepage     & 55  & (14\%)      & 65  & (17\%)  & 28  & (7\%)   \\
specification         & 21  & (5\%)       & 33  & (9\%)   & 32  & (8\%)   \\
organization homepage      & 16  & (4\%)       & 41  & (11\%)  & 24  & (6\%)   \\
other                 & 5   & (1\%)       & 23  & (6\%)   & 45  & (12\%)  \\
tutorial or article   & 16  & (4\%)       & 21  & (5\%)   & 31  & (8\%)   \\
API documentation     & 14  & (4\%)       & 20  & (5\%)   & 10  & (3\%)   \\
blog post             & 1   & (0\%)       & 10  & (3\%)   & 22  & (6\%)   \\
application           & 0   & (0\%)       & 11  & (3\%)   & 13  & (3\%)   \\
bug report            & 9   & (2\%)       & 10  & (3\%)   & 3   & (1\%)   \\
research paper        & 0   & (0\%)       & 9   & (2\%)   & 13  & (3\%)   \\
personal homepage     & 4   & (1\%)       & 8   & (2\%)   & 4   & (1\%)   \\
code                  & 6   & (2\%)       & 2   & (1\%)   & 5   & (1\%)   \\
forum thread          & 0   & (0\%)       & 5   & (1\%)   & 6   & (2\%)   \\
book content          & 0   & (0\%)       & 0   & (0\%)   & 2   & (1\%)   \\
GitHub profile        & 1   & (0\%)       & 0   & (0\%)   & 0   & (0\%)   \\
Stack Overflow thread & 1   & (0\%)       & 0   & (0\%)   & 0   & (0\%)   \\
Q\&A thread           & 0   & (0\%)       & 0   & (0\%)   & 1   & (0\%)   \\
\midrule
\textbf{sum}                   & \textbf{384} & \textbf{(100\%)}     & \textbf{384} & \textbf{(100\%)} & \textbf{378} & \textbf{(100\%)} \\
\bottomrule
\end{tabular}
\end{table}


\textbf{Taxonomy of link targets.}
Table~\ref{tab:rq2results} shows the result of our qualitative analysis. For commonly-linked domains, license is the most common type of link target, accounting for more than half of the links in our sample, followed by software homepages, i.e., the main web presence of a library or software project. For domains that are linked sometimes from source code comments, the most common type of link target was 404, a non-existing link target. This is a first indicator of the decay of links in source code comments, which we will analyze in detail in the next sections. Software homepages are also prevalent, as are organization homepages, both accounting for more than 10\% of all links in our sample. Finally, for links from domains which are rarely linked, the problem of decay is even more serious, affecting 37\% of the links in this sample. In other words, we can conclude with a 95\% confidence that between 32 and 42\% of all links to domains which are rarely linked from source code comments are dead or inaccessible. The prevalence of the code ``other'' in the results for links to rarely linked domains is an indicator of the diversity of links present in source code comments.

\begin{tcolorbox}
\textbf{Summary}: We identified more than a dozen different kinds of link targets, with dead links, licenses, and software homepages being the most prevalent. Dead links are particularly common for rarely linked domains.
\end{tcolorbox}

\subsection{Link Purpose (RQ3)}
To understand the purpose of links referenced in source code comments (\textbf{RQ3}) and similar to (\textbf{RQ2}), we again employed a qualitative analysis of our statistically representative and stratified sample of 1,146 links, only this time focusing on the origin of a link (in a source comment) rather than the target of the link. 
We used the same iterative approach to design a coding guide, and validated the coding guide by having the four authors code 30 links independently, this time leading to a kappa agreement of 0.70 which indicates ``substantial'' agreement~\cite{viera2005understanding}. 
The somewhat lower agreement can be explained by the need to extrapolate the purpose of a link from its context in the source code alone, without being able to interview the contributor who added the link. 

The following list shows all 8 codes that emerged from our analysis for link purpose, along with a short description which was available in the coding guide. The coding guide was informed by work on source code comments (e.g.,~\cite{storey2008todo}), self-admitted technical debt (e.g.,~\cite{potdar2014exploratory}), and attribution (e.g.,~\cite{baltes2018usage}).

\begin{itemize}
\item \textit{metadata}: the link relates to the author of the source code, a related organization, or the license
\item \textit{source/attribution}: the comment explicitly indicates that the link is a source of some aspect of the source code (e.g., algorithm)
\item \textit{source code context}: the link adds additional information to the source code (use this code for things that do not obviously fit into any of the previous)
\item \textit{see-also}: the comment explicitly indicates that the link points to additional reading material (usually accompanied by a phrase such as ``see also'').
\item \textit{commented-out source code}: the link is part of the source code, e.g., as a parameter value, but has been commented out
\item \textit{link-only}: the comment only contains the link
\item \textit{self-admitted technical debt}: bug-related, like workaround, under development, and so on
\item \textit{@see}: the link is accompanied by ``\texttt{@see}'', but no further explanation
\end{itemize}

Note that our coding guide required the indicators of \textit{see-also} and \textit{source/attribution} to be explicit, thus reducing the guesswork required as part of the qualitative analysis.

\begin{table}
\centering
\caption{Frequency of link purposes in our sample}
\label{tab:rq3results}
\begin{tabular}{l|r@{}rr@{}rr@{}r}
\toprule
 & \multicolumn{2}{c}{\textbf{common}} & \multicolumn{2}{c}{\textbf{sometimes}} & \multicolumn{2}{c}{\textbf{rare}} \\
\midrule
metadata                     & 288 & (75\%)  & 131 & (34\%)  & 43  & (11\%)  \\
source/attribution           & 27  & (7\%)   & 62  & (16\%)  & 75  & (20\%)  \\
source code context          & 18  & (5\%)   & 60  & (16\%)  & 80  & (21\%)  \\
see-also                     & 28  & (7\%)   & 59  & (15\%)  & 51  & (13\%)  \\
commented-out source code    & 1   & (0\%)   & 17  & (4\%)   & 70  & (19\%)  \\
link-only                    & 6   & (2\%)   & 24  & (6\%)   & 40  & (11\%)  \\
self-admitted technical debt & 11  & (3\%)   & 16  & (4\%)   & 13  & (3\%)   \\
@see                         & 5   & (1\%)   & 15  & (4\%)   & 6   & (2\%)   \\
\midrule
\textbf{sum}                   & \textbf{384} & \textbf{(100\%)}     & \textbf{384} & \textbf{(100\%)} & \textbf{378} & \textbf{(100\%)} \\
\bottomrule
\end{tabular}
\end{table}


\textbf{Taxonomy of link purpose.}
Table~\ref{tab:rq3results} shows the results of the qualitative analysis. For links to commonly linked domains, providing metadata, e.g., in the forum of licenses or author information, is by far the most common purpose of a link, covering three quarters of the links in our sample. For links to domains which are only sometimes linked, metadata only accounts for one third of the data, followed by links included for the purpose of attribution, providing context, or see-also information. The results for links to rarely linked domains are even more diverse: we can see from the table that these links are used for context, attribution, and as part of the source code functionality (albeit commented out), to name the top three. Six of the eight codes account for at least 10\% of the links in this part of our sample.

\begin{table}
\centering
\caption{Associations between link target type and link purpose}
\label{tab:arules}
\begin{tabular}{llclrr}
\toprule
\textbf{strata} & \multicolumn{3}{c}{\textbf{association rule}} & \textbf{conf.} & \textbf{supp.} \\
\midrule
common    & license                      & $\Rightarrow$ & metadata                     & 1.00 & 208 \\
common    & metadata                     & $\Rightarrow$ & license                      & 0.72 & 208 \\
common    & software homepage            & $\Rightarrow$ & metadata                     & 0.75 & 41  \\
common    & organization homepage             & $\Rightarrow$ & metadata                     & 0.88 & 14  \\
common    & bug report                   & $\Rightarrow$ & satd                         & 0.78 & 7   \\
common    & satd                         & $\Rightarrow$ & bug report                   & 0.64 & 7   \\
common    & personal homepage            & $\Rightarrow$ & metadata                     & 1.00 & 4   \\
\midrule
sometimes & software homepage            & $\Rightarrow$ & metadata                     & 0.65 & 42  \\
sometimes & organization homepage             & $\Rightarrow$ & metadata                     & 0.80 & 33  \\
sometimes & personal homepage            & $\Rightarrow$ & metadata                     & 1.00 & 8   \\
sometimes & license                      & $\Rightarrow$ & metadata                     & 1.00 & 4   \\
\midrule
rare      & personal homepage            & $\Rightarrow$ & metadata                     & 1.00 & 4   \\
\bottomrule
\end{tabular}
\end{table}

\textbf{Matching link target with purpose.}
Based on the qualitative analysis conducted for answering \textbf{RQ2} and \textbf{RQ3} about the targets and purposes of links in source code comments, we are now able to investigate the relationships between the different types of link targets and the different purposes which emerged from our qualitative analysis. To do so, we applied association rule learning using the \textit{apriori} algorithm~\cite{AS1994} as implemented in the R package \texttt{arules}\footnote{\url{https://cran.r-project.org/web/packages/arules/index.html}} to our data, treating each link as a transaction containing two items: its target type and its purpose. We used 4 as threshold for support and 0.7 as threshold for confidence, i.e., all rules that we extracted are supported by at least four data points and we have at least a 70\% confidence that the left hand side of the rule implies the right hand side.

Table~\ref{tab:arules} shows the association rules extracted from our data with these settings, separately for each stratum in our sample. Unsurprisingly, the link target type \textit{license} and the purpose of providing \textit{metadata} are tightly connected, in particular for links referring to commonly linked domains. In fact, all links to licenses were found to have been included for the reason of providing metadata, and 72\% of the metadata is license information. Links to software, organization, and personal homepages are also associated with metadata, across all strata. Although with a relatively low support of seven instances, it is also interesting to note the tight coupling of the link target type \textit{bug report} and the purpose of admitting technical debt.

\begin{tcolorbox}
\textbf{Summary}: We identified different purposes for the inclusion of links in source code comments, with providing metadata and attribution being the most common. Links are also included for background information, to provide context, or to admit technical debt. In some cases, the link is part of source code which has been commented out.
\end{tcolorbox}

\subsection{Link Evolution (RQ4)}
\label{ssec:evol}
To understand how links evolve (\textbf{RQ4}), we investigated the revision histories of repositories in the samples from (\textbf{RQ2}).
For each sample link, we searched an old version of the link that has been revised by a commit that introduced the link.
We extracted such a commit introducing a link by using the git log command (\texttt{-S} option with tracking file renaming).
We searched http(s) links removed from the code location where the sample link has been added.
We identified 88 revised links out of 1,146 samples, including 24 (6.3\%) in common, 31 (8.1\%) in sometimes, and 33 (8.7\%) in rare.
Less than 10\% of the links had been revised in each strata, that is, most of the links have never been updated.

We manually analyzed the old and new paths of the links and identified the following evolution types:
\begin{itemize}
\item \textit{license replacement}: a new link refers to a new software license. For example, a link to GNU GPL has been replaced with a link to the Apache License.
\item \textit{organization update}: a project or an organization changed its name or website.  For example, a project that acquired their own domain updated links to their project website.
\item \textit{change to https}: a new link uses HTTPS instead of HTTP for the same location as the previous link.
\item \textit{content move}: a new link refers to a slightly different location (e.g. the same path on a different server, the same document name on a different wiki), which is likely the same content.  
\item \textit{content update}: a new link refers to different content from the previous link, but the new content is likely updated.  
For example, the Apache Jackrabbit project replaced a link pointing to a draft version of a document\footnote{\url{http://greenbytes.de/tech/webdav/draft-ietf-webdav-bind-21.html}} with a link to an RFC version.\footnote{\url{http://greenbytes.de/tech/webdav/rfc5842.html}}
\item \textit{content change}: a new link refers to relevant but different content from the previous link.
For example, the Pi4J project replaced a link related to the usage of a serial port of Raspberry Pie\footnote{\url{http://www.irrational.net/2012/04/19/using-the-raspberry-pis-serial-port/}} with another similar document.\footnote{\url{https://www.cube-controls.com/2015/11/02/disable-serial-port-terminal-output-on-raspbian/}}
\item \textit{other}: we could not identify types for some links whose contents are no longer available.  
It should be noted that the contents for 20 updated links are 404 Not Found.
\end{itemize}

\begin{table}[t]
\caption{Link Evolution Types}
\label{tab:rq4summary}
\begin{center}
\begin{tabular}{l|r r r r r r}
\toprule
 & \multicolumn{2}{c}{\textbf{common}} & \multicolumn{2}{c}{\textbf{sometimes}} & \multicolumn{2}{c}{\textbf{rare}} \\  \midrule
license & 12 & (50\%) & 1 & (3\%) & 1 & (3\%) \\ 
organization & 7 & (29\%)  & 10 & (32\%) & 3 & (9\%) \\
https & 2 & (8\%) & 1 & (3\%) & 5 & (15\%) \\
content move & 0 & (0\%) & 2 & (6\%) & 6 & (18\%) \\
content update & 1 & (4\%) & 3 & (10\%) & 4 & (12\%) \\
content change & 0 & (0\%) & 6 & (19\%)  & 2  & (6\%) \\
other & 2 & (8\%) & 8 & (26\%) & 12 & (36\%) \\ 
\midrule
\textbf{sum} & \textbf{24} & \textbf{(100\%)} & \textbf{31} & \textbf{(100\%)} & \textbf{33} & \textbf{(100\%)} \\  
\bottomrule
\end{tabular}
\end{center}
\end{table}


\textbf{Reasons for link evolution.}
Table~\ref{tab:rq4summary} shows the numbers of link evolution in the three strata.
For commonly-linked domains, license replacement and updating organizational information account for about 80\% of link revisions. For domains sometimes linked, organization update is the most common, followed by other and content change. For rarely-linked domains, links are revised for content move, content update, and change to https.

\begin{tcolorbox}
\textbf{Summary}: Links are rarely updated (less than 9\%). Common modifications are updating licenses and organization homepages.
\end{tcolorbox}

\subsection{Link Target Evolution (RQ5)}
\label{ssec:shortevol}

After understanding the evolution of links, our next research question \textbf{(RQ5)} asks about the evolution of their targets. To investigate whether link targets referenced in source code comments evolve, we attempted to download all link targets in our sample of 1,146 links using the \texttt{curl} command with a timeout of 60 seconds. 
As already discussed as part of \textbf{(RQ2)}, not all link targets are available. 
We were able to download a total of 1,034 link targets (90\%). 
We then repeated the same download process exactly ten days later, to see how many of the link targets had changed within this short time frame and what kind of changes had happened.

\begin{table}
\centering
\caption{Link target evolution within a 10-day timeframe}
\label{tab:target_evol}
\begin{tabular}{lr}
\toprule
\textbf{evolution} & \textbf{\# of links} \\
\midrule
no content returned on both dates             & 112 \\
no changes between the two dates              & 879 \\
content not available anymore                 & 6   \\
\midrule
auto-generated changes (e.g., \# of visitors) & 125 \\
content changed                               & 7   \\
``latest'' updated                            & 7   \\
design change (e.g., new banner added)        & 6   \\
different error (e.g., 502 $\Rightarrow$ 301) & 3   \\
page title changed                            & 1   \\
\midrule
\textbf{sum}                                           & \textbf{1,146} \\
\bottomrule
\end{tabular}
\end{table}

\textbf{Changes to the link target.}
Table~\ref{tab:target_evol} summarizes the results of this analysis: out of the 1,034 link targets for which \texttt{curl} returned a result, 879 (85\%) had not changed at all in the ten-day time frame (the downloaded content was exactly the same, as per the Windows file compare tool \texttt{fc}). We manually analyzed the 155 cases in which the content had changed by opening both versions in a web browser and conducting a visual comparison. The majority of the changes in the remaining 15\% can be attributed to automatically generated changes, such as the display of a visitor count or the current date in a footer.

However, a non-negligible number of link targets underwent more significant changes in the ten-day time window: For six links for which we were able to retrieve data on the first download date, there was no content available anymore ten days later. For three links which had displayed an error message when we first attempted to download their content, the specific error message changed. Some link targets changed their website design, and for a few links, the content changed. For example, the download page of TaskWarrior\footnote{\url{https://taskwarrior.org/download/}} included the following notice when we first downloaded its content: ``(For those of you wishing to build task from source on Cygwin, you will need some components installed (make, g++/clang, GnuTLS, libuuid, libreadline), but don't forget - task is a standard part of the Cygwin distribution, so you do not need to build from source, unless you want the latest development snapshot).'' Ten days later, this notice was replaced with: ``(Please note, that Cygwin is not supported anymore. Please use the Windows Subsystem for Linux to use Taskwarrior on Windows).'' We argue that this kind of change is relevant to software developers.

\textbf{Stack Overflow case study.}
To investigate this phenomenon in more detail, we conducted a case study with the subset of links pointing to Stack Overflow. As seen in Section~\ref{ssec:prevalence}, \texttt{stackoverflow.com} is the second most referenced domain.

In all 9,654,702 obtained links, there are 32,197 links belonging to \texttt{stackoverflow.com}. Among those Stack Overflow links, there are varieties of expressions: an abbreviated path to an answer (\texttt{/a/(answer id)}), an abbreviated path to a question (\texttt{/q/(question id)}), and a full path to a question (\texttt{/questions/(question id)/(title)}). Older links start with `\texttt{http://}' and newer links start with `\texttt{https://}'.
For each Stack Overflow link, we identified the timestamp of when the link was added to a repository by using the same git log command (\texttt{-S} option with tracking file renaming) used in Section~\ref{ssec:evol}. 
For duplicate links, we consider only the oldest timestamp. Consequently, we obtained a list of 11,464 distinct links with their timestamps.

We then made use of the SOTorrent dataset~\cite{baltes2018sotorrent} to investigate the extent to which Stack Overflow content had changed since the link to the question or answer had been added to a source code comment in a Git repository. We created a statistically representative sample of 372 links from the population of all unique links to Stack Overflow content in our dataset, and we queried SOTorrent to determine the following metrics for each link:

\begin{itemize}
    \item the number of text edits on any post (i.e., question or answer) in the same thread,
    \item the number of new comments on any post (i.e., question or answer) in the same thread,
    \item the number of new answers in the same thread, and
    \item the number of edits to the thread title.
\end{itemize}

\begin{figure}
\centering
\includegraphics[width=\linewidth]{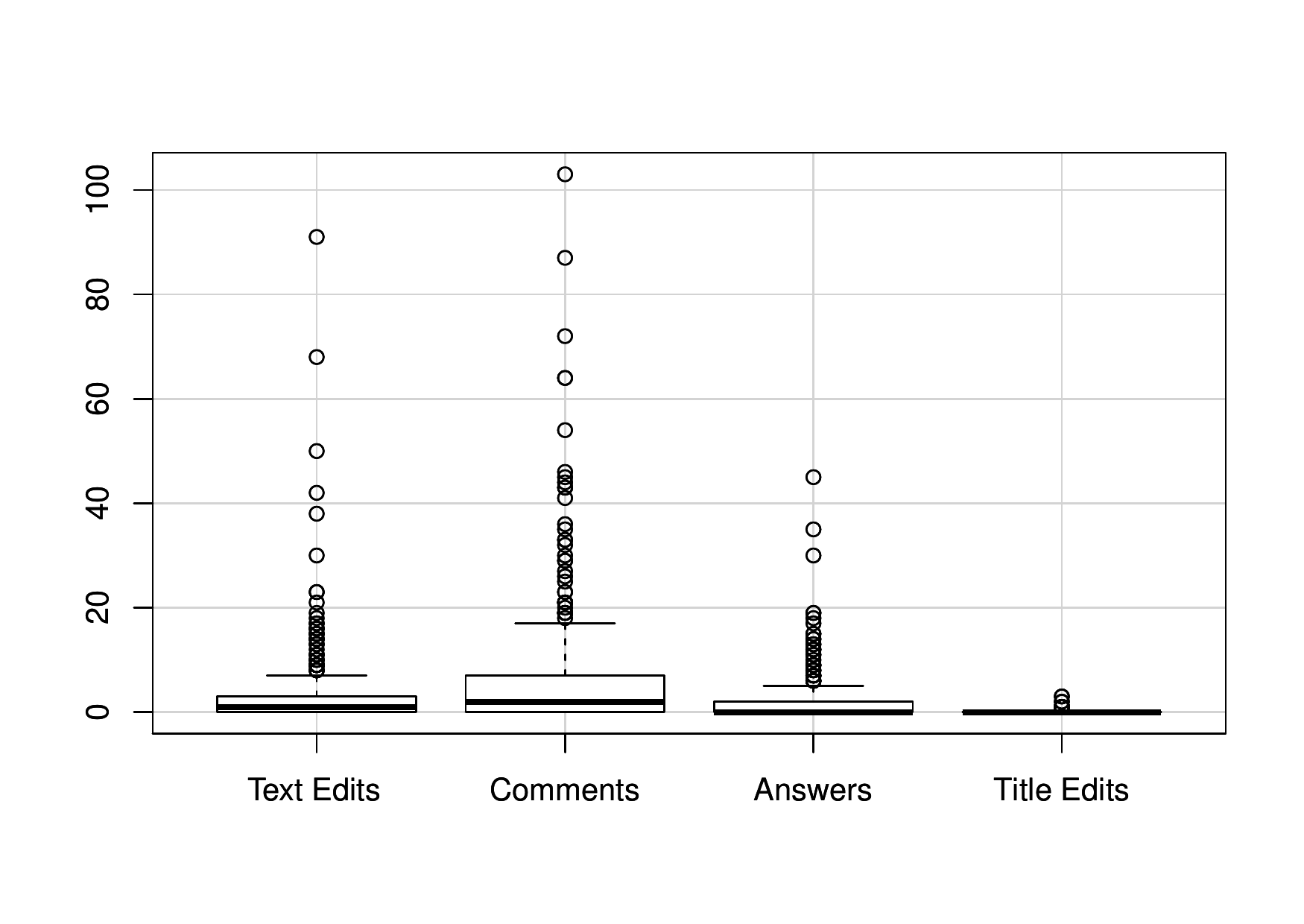}
\caption{Number of changes to the linked Stack Overflow threads after they were first referenced}
\label{fig:RQ6}
\end{figure}

\textbf{Thread updates.}
Figure~\ref{fig:RQ6} shows the results of this analysis. More than half of all Stack Overflow threads had at least one change made to the text of a question or answer in the same thread (median: 1, third quartile: 3) after they were committed to a Git repository as part of a source code comment, and more than half of these links attracted at least one new comment in the meantime (median: 2, third quartile: 7). While the number of new answers to a thread was zero in the median case, a quarter of the Stack Overflow threads attracted at least 2 new answers after the link was added in a source code comment (median: 0, third quartile: 2). In total, only 91 (24\%) of the 372 Stack Overflow threads in our sample did not undergo any changes after they were added to a Git repository.

The most extreme example in our sample---a Stack Overflow thread titled ``How do you split a list into evenly sized chunks?''\footnote{\url{https://stackoverflow.com/q/312443}}---had 45 new answers added (out of a total of 58) and attracted 103 new comments, 50 edits, and one change to the thread title after being first referenced in a source code comment in early September 2010.

\begin{tcolorbox}
\textbf{Summary}: We found that even within a short ten-day time window, a non-negligible portion of link targets referenced in source code comments evolve, in some cases adding or modifying pertinent information. In our case study on links pointing to Stack Overflow, we found that more than three quarters of all Stack Overflow threads linked in source code comments attracted at least one change (edit, new answer, or new comment) after being first referenced in a source code comment.
\end{tcolorbox}

\subsection{Link Decay (RQ6)}
\label{ssec:decay}

Among the obtained 9,654,702 links, there are 382,650 distinct links.
To investigate the amount of dead links in source code comments (\textbf{RQ6}),
we accessed all Web contents from the 382,650 unique links by using the Perl module \texttt{LWP}.\footnote{We used \texttt{LWP::UserAgent} and \texttt{LWP::RobotUA}.}


\textbf{Link retrieval responses.}
Table~\ref{tab:status} shows obtained HTTP Status Codes for all unique links. 
More than 80\% of the links were successfully reachable including redirecting to new locations.
Although some HTTP status codes do not clearly mean links are dead, we found that
about 9\% of the links are not available now (404 Not Found). Among these 404 links, we saw
some patterns: Internet host lost, Web content move/deletion, fake URL (for showing examples of parameters or data structures), and link typo. 
The domain with the largest number of 404s is \texttt{github.com}, with 3,346 links no longer available or pointing to private repositories.

\begin{table}
\centering
\caption{HTTP Status Codes for All Unique Links}
\label{tab:status}
\begin{tabular}{lrr}
\toprule
\textbf{status} & \textbf{\# links} & \textbf{(\%)} \\
\midrule
2xx success & 310,592 & (81.2\%) \\
404 not found & 34,689 & (9.1\%) \\
500 internal server error & 18,955 & (5.0\%) \\
405 method not allowed & 10,104 & (2,6\%) \\
403 forbidden & 4,068 & (1.0\%) \\
others & 4,242 & (1.1\%) \\
\midrule
\textbf{sum} & \textbf{382,650} & \textbf{(100\%)} \\
\bottomrule
\end{tabular}
\end{table}

\begin{tcolorbox}
\textbf{Summary}: We found that 9.1\% of the link targets are not available (404 Not Found), considering all unique links.
\end{tcolorbox}

\subsection{Fixing Dead Links (RQ7)}
To fix dead links (\textbf{RQ7}), we collected fixable dead links and submitted pull requests to fix.
We select dead links that are not metadata (need multiple files to be fixed) nor commented-out source code.
Personal blog articles were avoided because they tend to be no longer available.
Consequently we obtained 14 dead links to API documentation, research papers, and so on.
After checking the original content in \texttt{Wayback Machine}\footnote{\url{https://web.archive.org/}}, we manually investigated new links by searching specific keywords in the original content.
%
Our fixing process included first forking a personal copy of the project, fixing the link, and then later submission of a pull request to the project.



\textbf{Pull request results}.
Developers showed they cared about dead links by accepting to all nine pull requests.
\footnote{\url{https://github.com/mockingbirdnest/Principia/pull/1905}}
\footnote{\url{https://github.com/sveawebpay/php-integration/pull/82}}
\footnote{\url{https://github.com/shirasagi/shirasagi/pull/2289}}
\footnote{\url{https://github.com/onepercentclub/bluebottle/pull/3372}}
\footnote{\url{https://github.com/cms-sw/cmssw/pull/24370}}
\footnote{\url{https://github.com/BrowserSync/browser-sync/pull/1593}}
\footnote{\url{https://github.com/ShipSoft/FairShip/pull/133}}
\footnote{\url{https://github.com/BlackToppStudios/Mezzanine/pull/182}}
\footnote{\url{https://github.com/onepercentclub/bluebottle/pull/3372}}

Since the link itself is a comment, we speculate that it has almost no conflicts with existing code, so our pull requests are likely to pass all tests and to be merged immediately.
Developers responded with comments such as \textit{``LGTM (looks good to me)''} and \textit{``Thanks for spotting the broken link''}.

Overall, the responses from developers provide sufficient motivation for tool support to assist with fixing broken links.
We argue that such comments indicate that developers are concerned with keeping their links alive.

\begin{tcolorbox}
\textbf{Summary}: Developers generally responded positively to the request to fix dead links. 
All nine responsive projects accepted our pull requests to fix dead links.
\end{tcolorbox}


\section{Recommendations}
\label{sec:recom}

Our findings can be summarized into recommendations for developers and researchers.

Recommendations for software developers including links in source code comments are:
\begin{itemize}
    \item \textit{Try referencing permanent links}, as it is reported that more than 30\% of links will not work after a 4 year period~\cite{Koehler:2002:WPC:506072.506080}.
    Referencing research papers with DOI is preferable instead of researchers' personal Web pages.
    Explicitly mentioning tags or commit hashes to referenced code in GitHub would be recommended, as software structure can be changed (we found many dead links to GitHub in Section~\ref{ssec:decay}).
    \item \textit{Check link targets for new information on a regular basis}, as referenced external resources can be considered to be software documentation to support comprehension and maintenance activities. In addition, link target updates can be triggers of improving and updating code (as seen in Section~\ref{ssec:shortevol}). 
\end{itemize}

We can also consider future work with the following possible challenges.
\begin{itemize}
    \item \textit{Further understanding of external sources}.
    We found many sources as shown in Figure~\ref{fig:RQ1c} and Table~\ref{tab:rq2results}.
    Although some sources have been already studied, for example,
    licenses~\cite{German:2010:SMA:1858996.1859088}, self-admitted technical debt~\cite{potdar2014exploratory}, and Stack Overflow~\cite{Treude:2011:PAA:1985793.1985907},
    other sources have not been well-studied with regard to their impact and influence on software development, such as research papers and Wikipedia articles.
    \item \textit{Further studies of source code comments} to understand how knowledge (related to knowledge-based theory~\cite{ZAHEDI201636} and human capital~\cite{WOHLIN2015229,DBLP:journals/corr/abs-1805-03844}) is summarized and shared via source code comments. Further analyses of source code comment contents~\cite{Pascarella:2017:CCC:3104188.3104217} would be required.
    \item \textit{Tool support for external source referencing, tracking, and updating}. Although we recommend developers to maintain links and associated code, it is not always possible. Tools or systems to help developers fix link issues and maintain code automatically could be practically useful.
\end{itemize}

\section{Threats to Validity}
\label{sec:ttv}

Threats to the \textit{construct validity} exist in our approach to link identification. Since we identified links per line in source code comments, links located across multiple lines cannot be extracted. Note that we did not encounter any such multiple-line links in our representative sample of 1,146 links. Hence we consider that the impact of incorrect link identification because of multiple-line links is small.

Threats to the \textit{external validity} exist in our repository preparation. Although we analyzed a large amount of repositories on GitHub, we cannot generalize our findings to industry nor open source projects in general; some open source repositories are hosted outside of GitHub, e.g., on GitLab or private servers.

To mitigate threats to \textit{reliability}, we prepared an online appendix of our 9,654,702 links with associated information (see Section~\ref{ssec:appendix}).


\section{Related Work}
\label{sec:rw}
We have discussed complementary work throughout the paper in the relevant sections; here, we discuss literature related to web resources and source code comments.

One of the most related studies is the one by
Xia~et al. \cite{XiaEMSE2017}. They investigated what developers search for on the Web, and found that developers search for explanations of unknown terminology, explanations for exceptions/error messages (e.g., HTTP 404), reusable code snippets, solutions to common programming bugs, and suitable third-party libraries/services.
Furthermore, they found that searching for solutions to performance bugs, solutions to multi-threading bugs, public datasets to test newly developed algorithms or systems, reusable code snippets, best industrial practices, database optimization solutions, solutions to security bugs, and solutions to software configuration bugs are the most difficult search tasks that developers consider.

Many researchers have made use of code comments in their work.
Tan et al. \cite{Tan:2007:IBB:1294261.1294276} automatically identify bugs by analyzing inconsistencies between code and comments.
Ratol and Robillard \cite{Ratol2017} used code comments to assist refactoring activities.
Wong~et al. \cite{Wong:ase2013} used code comments to map source code and Stack Overflow content.
German~et al. \cite{German:2010:SMA:1858996.1859088} developed the ninka tool that automatically identifies a software license in code comments.  
Goldman and Miller \cite{GOLDMAN2008VL} developed the tool CodeTrail, that demonstrates how the developer's use of web resources can be improved by connecting the Eclipse integrated development environment (IDE) and the Firefox web browser.


Self-admitted technical debt is a commenting activity that has been well-studied in recent years~\cite{potdar2014exploratory}.
Maldonado et al. \cite{8094425} and Zampetti et al. \cite{Zampetti:2018:STD:3196398.3196423} studied the removal of self-admitted technical debt based on the modification of comments.
Our finding of referencing bug reports for self-admitted technical debt could be another opportunity to study development activities around technical debt.

There are also studies which analyze link sharing occurring in other software artifacts.
Gomez~et al. \cite{Gomez:2013} investigated link sharing on Stack Overflow to gain insights into how software developers discover and disseminate innovations.
Rath~et al. \cite{Rath2018} investigated links to issue tracking systems in commit comments.
They reported that developers often do not provide external links to issues.  They evaluated several methods to automatically recover links by searching issues related to a given commit.
Alqahtani~et al. \cite{Alqahtani2017} proposed a tool to automatically link dependent components in a system to online resources for analyzing their vulnerabilities.
Chen~et al. \cite{Chen2015} proposed a tool to link problematic source code to relevant Stack 
Overflow questions using similarity of source code fragments.

Traceability links between source code and documents 
is another related research topic.
Scanniello~et al. \cite{Scanniello2018} reported that developers can understand source code effectively if they can refer to design models including source code element names. 
Their observation has been obtained through a controlled experiment of program comprehension tasks with UML models produced in a requirements engineering phase and a design phase.
Antoniol~et al. \cite{Antoniol2000} proposed a method to identify links between source files and design documents because developers may update source file names without updating related documents.
Their method uses similarity of attribute names of a class to identify its original class definition in design documents.
Rahimi~et al. \cite{Rahimi2017} proposed a rule-based method to update links between source files and requirements documents.
Their method recognizes a change scenario from semantic differences of source code and then updates links according to a rule corresponding to the change scenario.
Those methods would be effective to automatically update traceability links. Similar tool support for external source referencing is a future direction of our research.

However, none of the related work provides a comprehensive study of the role of links in source code comments, which is the goal of this paper.

\section{Conclusion}
\label{sec:cfw}



To understand purposes, evolution, and decay of links in source code comments, we conducted
(i) a quantitative study of 9,654,702 links from source code comments in 25,925 Git repositories to establish the prevalence of links in source code comments;
(ii) a qualitative study of a stratified sample of 1,146 links to determine the kinds of link targets and purposes for including links present in our dataset;
(iii) a quantitative and qualitative study to investigate the evolution of links in source code comments and their targets; and
(iv) a quantitative study to determine the extent to which links in source code comments are affected by link decay.

Our work has shown that links in source code comments indeed suffer from decay, from insufficient versioning (when link targets evolve), and from lack of bidirectional traceability (which could help avoid decay). Based on this work which has established the prevalence of links in source code comments, their multiple purposes and targets, issues of decay, and practical needs of fixing dead links, there are many open avenues for future work: understanding the role of external sources for software development, further studies of source code comments, and tool support for external source referencing, to name a few. 

\section*{Acknowledgment}

We thank the respondents to our pull requests for their availability and Sebastian Baltes for his support in querying the SOTorrent dataset.
This work has been supported by JSPS KAKENHI Grant Numbers JP16H05857, JP17H00731, and JP18KT0013 as well as the Australian Research Council's Discovery Early Career Researcher Award (DECRA) funding scheme (DE180100153).


\end{document}